\documentclass{Interspeech}

\usepackage{color}

\newcommand{\eedit}[1]{\textcolor{red}{#1}}
\newcommand{\hedit}[1]{\textcolor{blue}{#1}}

\interspeechcameraready

\title{RapFlow-TTS: Rapid and High-Fidelity Text-to-Speech with \\ Improved Consistency Flow Matching}

\author[affiliation={1,2,*}]{Hyun Joon}{Park}
\author[affiliation={1}]{Jeongmin}{Liu}
\author[affiliation={2}]{Jin Sob}{Kim}
\author[affiliation={2}]{Jeong Yeol}{Yang}
\author[affiliation={2}]{Sung Won}{Han}
\author[affiliation={1,3}]{Eunwoo}{Song}

\affiliation{}{NAVER Cloud}{Republic of Korea}
\affiliation{School of Industrial and Management Engineering}{Korea University}{Republic of Korea}
\affiliation{Artificial Intelligence Institute}{Seoul National University}{Republic of Korea}

\email{
    winddori2002@korea.ac.kr
}


\usepackage{comment}

\begin{document}

\maketitle



\begin{abstract}

We introduce RapFlow-TTS, a rapid and high-fidelity TTS acoustic model that leverages velocity consistency constraints in flow matching (FM) training. Although ordinary differential equation (ODE)-based TTS generation achieves natural-quality speech, it typically requires a large number of generation steps, resulting in a trade-off between quality and inference speed. To address this challenge, RapFlow-TTS enforces consistency in the velocity field along the FM-straightened ODE trajectory, enabling consistent synthetic quality with fewer generation steps. Additionally, we introduce techniques such as time interval scheduling and adversarial learning to further enhance the quality of the few-step synthesis. Experimental results show that RapFlow-TTS achieves high-fidelity speech synthesis with a 5- and 10-fold reduction in synthesis steps than the conventional FM- and score-based approaches, respectively.


\end{abstract}

\section{Introduction}
\label{sec:intro}


\renewcommand{\thefootnote}{}
\footnotetext{* Work performed as an intern in the Voice team, NAVER Cloud.}
\renewcommand{\thefootnote}{\arabic{footnote}} 

Text-to-speech (TTS), also known as speech synthesis, aims to synthesize high-fidelity speech, given an input text \cite{wang2017tacotron, shen2018natural, kim2020glow}. Among the various generative modeling approaches \cite{ren2019fastspeech, miao2020flow, li2019neural, ren2020fastspeech, jeong2021diff}, ordinary differential equations (ODE)-based models \cite{popov2021grad, park2024dex} have become strong solutions for outstanding TTS. 
One representative method involves diffusion models using stochastic differential equations (SDE) \cite{song2020score, karras2022elucidating}, e.g., Grad-TTS \cite{popov2021grad}.
It trains a score network with an SDE-based diffusion process and solves the probability flow ODE for speech synthesis.
However, ODE trajectories obtained by diffusion models are complicated \cite{lipman2022flow, tong2023improving}, requiring numerous steps to generate high-quality speech \cite{guo2024voiceflow}. 
Since TTS functions as a user interaction channel, the slow inference speed caused by numerous diffusion steps is a major drawback of diffusion-based TTS.

To resolve this limitation, researchers investigated improved ODE-based models \cite{lipman2022flow, tong2023improving, liu2022flow, song2023consistency} for TTS. 
For example, Comospeech \cite{ye2023comospeech} applied consistency distillation \cite{song2023consistency} to transfer the knowledge of a diffusion teacher model to a consistent student model. 
As the consistency constraint maps any point on the ODE trajectory to the endpoint, i.e., the target data, the student model can consistently produce high-quality speech regardless of the number of diffusion steps.
However, the complexity of diffusion-based ODE trajectories limited the effective construction of consistency models along those trajectories.

Meanwhile, flow matching (FM), which learns the vector field for the ODE and probability path, is another approach for improving ODE-based models.
Using an optimal transport plan that makes the probability path linear in time \cite{lipman2022flow, tong2023improving}, the ODE trajectory tends to be straight, making the few-step performance of FM superior to that of the diffusion models. 
In TTS frameworks, Matcha-TTS \cite{mehta2024matcha} and VoiceFlow \cite{guo2024voiceflow} adopted the FM-based straight ODE trajectory, demonstrating high-quality speech synthesis with even fewer generation steps than conventional diffusion-based models.
However, completely generalizing the real distribution to a straight ODE trajectory is difficult; thus, the quality of the few-step synthesized speech has not yet reached that of the ground truth. 
In summary, previous ODE-based TTS approaches failed to simultaneously satisfy both aspects, i.e., inference speed and speech synthesis quality.

In this study, we present RapFlow-TTS, a rapid and high-fidelity TTS acoustic model that effectively addresses the limitations above. Inspired by advancements in image generation, we adopt consistency FM \cite{yang2024consistency}, a novel FM method that explicitly enforces self-consistency in the velocity field. By imposing consistency constraints on the velocity values, consistency FM directly defines straight flows from different time points to a common endpoint. This allows RapFlow-TTS to learn how to produce consistent outputs along the straight trajectory more effectively, unlike diffusion-based methods with complicated trajectories, yielding a more effective consistency model. Consequently, high-quality speech synthesis is achievable with only a few generation steps. To the best of our knowledge, RapFlow-TTS is the first TTS framework based on consistency FM.

Furthermore, we propose several improved training techniques to enhance the performance, including shared dropout, Huber loss, time-interval scheduling, and adversarial learning, none of which have been explored in the context of consistency FM.
Experimental results show that the proposed RapFlow-TTS can synthesize more natural speech in just 2 steps, reducing generation steps by 5 to 10 times compared to previous ODE-based methods.
Our code and demos are available here.\footnote{\url{https://tts-demo.github.io/rap.github.io/}}

\begin{figure*}[t]
  \centering
  \includegraphics[scale=0.58]{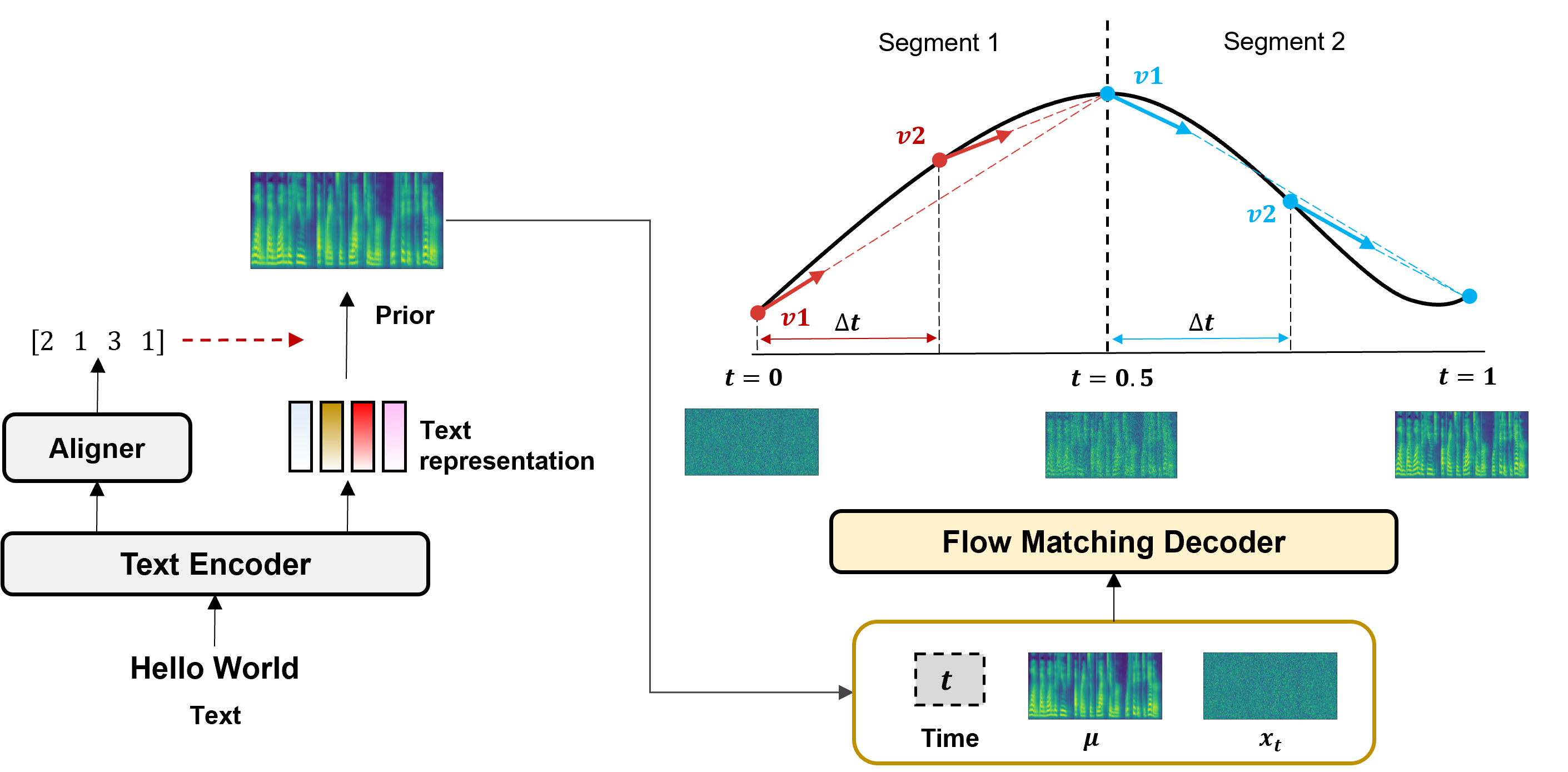} 
  \caption{Overview of RapFlow-TTS with multi-segment consistency flow matching.} 
  \label{fig:rapflow}
\end{figure*}

\section{Consistency Flow Matching}
\label{sec:cfm}

Here, we review the consistency FM \cite{yang2024consistency} method for training the RapFlow-TTS.
Given a time index $t\in[0,1]$, sampled from a uniform distribution, FM learns a ground truth vector field $u_{t}$ to build the probability path $p_{t}(x_{t})$ from the random noise $x_{0} \sim p_{0}$ to target data $x_{1} \sim p_{1}$.
The vector field is defined as an ODE for a flow $\phi_{t}$, where $\phi_{t}$ is a function that transforms data over time, as follows:
\begin{equation}\label{eq:ode}
  \begin{gathered}
    \frac{d}{dt}\phi_{t}(x)=u_{t}(\phi_{t}(x)); \ \  \phi_{0}(x)=x,
  \end{gathered}
\end{equation}
The solution of Eq. (\ref{eq:ode}), i.e., flow, represents the sampling path. For training FM models, a simple regression with the ground-truth vector field is applied as follows:
\begin{equation}\label{eq:fm}
  \begin{gathered}
    \mathcal{L}_{FM}=||v_{\theta}(t, x_{t}) - u(t, x_{t})||^{2}_{2},
  \end{gathered}
\end{equation}
where $v_{\theta}(t, x_{t})$ denotes the learnable vector field network.

In contrast to the FM, consistency FM trains the vector field by ensuring that any point on the trajectory reaches the same endpoint with the same velocity. Let the linearly interpolated noise sample at time $t$ be ${x}_{t}=tx_{1} + (1-t)x_{0}$. Its objective function is composed of a straight flow loss $\mathcal{L}_{sf}$ and a velocity consistency loss $\mathcal{L}_{vc}$ as follows:
\begin{equation}\label{eq:cfm}
  \begin{gathered}
    \mathcal{L}_{cfm}=\mathcal{L}_{sf}+\alpha\mathcal{L}_{vc},\\
    \mathcal{L}_{sf}=||f_{\theta}(t, x_{t}) - f_{\theta^{-}}(t+\Delta t, x_{t+\Delta t})||^{2}_{2},\\
    \mathcal{L}_{vc}=||v_{\theta}(t, x_{t}) - v_{\theta^{-}}(t+\Delta t, x_{t+\Delta t})||^{2}_{2},\\
    f_{\theta}(t,x_{t})=x_{t}+(1-t) \times v_{\theta}(t,x_{t}),
  \end{gathered}
\end{equation}
where $\theta^{-}$ denotes model parameters with a stop gradient; $\alpha$ denotes a loss weight; and $\Delta t$ denotes a time interval.
Since $f_{\theta}$ guides $x_{t}$ to the estimated endpoint of the trajectory, the straight flow loss constrains consistency from the trajectory viewpoint, ensuring straight flow. In contrast, the velocity consistency loss directly enforces the consistency of the vector field.
This allows the consistency FM to learn how to produce consistent outputs along the straightened trajectory, accurately estimating the target distribution even with a few-step generation.

\section{RapFlow-TTS}

\subsection{TTS with Consistency Flow Matching}
\label{ssec:cfmtts}

For RapFlow-TTS, we follow the network design of Matcha-TTS \cite{mehta2024matcha} thanks to its fast and lightweight properties. As shown in Figure \ref{fig:rapflow}, RapFlow-TTS consists of a text encoder, an aligner, and a flow matching decoder. The text encoder extracts the context representations from the input text, and the aligner maps these representations to prior mel-spectrograms $\mu$. The text encoder and aligner are trained using duration loss $\mathcal{L}_{dur}$ and prior loss $\mathcal{L}_{prior}$ based on the MAS algorithm as in \cite{kim2020glow, popov2021grad}.

Consistency FM trains the FM decoder to construct a probability path $p_{t}$ that transports the random noise $x_{0} \sim p_{0}$ to target mel-spectrogram's distribution $x_{1} \sim p_{1}$. We condition the consistency FM objective with the prior $\mu$ to synthesize speech corresponding to the given input text. 
We also adopt a multi-segment objective for flexible transportation between complex distributions \cite{yang2024consistency}.
In detail, the time range $t \in [0,1]$ is divided into $S$ segments of equal size, where each segment corresponds to the time range $[i/S, (i+1)/S]$ for $i=0,1,..., S-1$.
Then, the objective function of segmented consistency FM is expressed as follows:
\begin{equation}\label{eq:ms-cfm}
  \begin{gathered}
    \mathcal{L}_{cfm}=\mathcal{L}_{sf}+\alpha\mathcal{L}_{vc},\\
    \mathcal{L}_{sf}=||f^{i}_{\theta}(t, x_{t}, \mu) - f^{i}_{\theta^{-}}(t+\Delta t, x_{t+\Delta t}, \mu)||^{2}_{2}, \\
    \mathcal{L}_{vc}=||v^{i}_{\theta}(t, x_{t}, \mu) - v^{i}_{\theta^{-}}(t+\Delta t, x_{t+\Delta t}, \mu)||^{2}_{2}, \\
    f^{i}_{\theta}(t,x_{t}, \mu)=x_{t}+((i+1)/S-t) \times v^{i}_{\theta}(t,x_{t}, \mu),
  \end{gathered}
\end{equation}
Consequently, RapFlow-TTS learns straight flow with a consistent vector field in each segment, effectively representing the data distribution by a piecewise linear trajectory.

In practice, we apply a two-stage training strategy to effectively optimize the RapFlow-TTS model.
In the first stage (the first $N$ epochs), we aim to train RapFlow-TTS to have a straight flow by focusing on the trajectory viewpoint.
To achieve it, we optimize RapFlow-TTS using only the straight flow loss $\mathcal{L}_{sf}$ in Eq. (\ref{eq:ms-cfm}), slightly modifying the loss term as $||f^{i}_{\theta}(t, x_{t}, \mu)-x^{i}||^{2}_{2}$, where $x^{i}$ is the ground-truth endpoint at each segment, defined as $x^{i}=(i+1)/S \times x_{1} + (1-(i+1)/S)\times x_{0}$.
Because this loss guides the ground-truth endpoint on the trajectory, RapFlow-TTS can learn the vector field with a straight flow that represents the real data distribution. 

In the second stage (the following $N$ epochs), we train RapFlow-TTS to have a straight flow with a consistent vector field using the entire $\mathcal{L}_{cfm}$ in Eq. (\ref{eq:ms-cfm}). 
By training consistency FM on a straight ODE trajectory, RapFlow-TTS can learn consistency more easily than diffusion-based methods, constructing a more effective consistency model.
As a result, RapFlow-TTS has the properties of consistency and straight flow, thus synthesizing high-quality speech in fewer synthesis steps.

\subsection{Improved techniques for RapFlow-TTS}

We introduce several techniques to improve the consistency FM model for RapFlow-TTS.
All of these methods enhance few-step synthesis performance, and their effectiveness will be further analyzed in Section \ref{ssec:ablation}.

\noindent \textbf{Encoder Freeze}
We freeze the encoder and optimize only $\mathcal{L}_{cfm}$ for training efficiency in the second stage.
It also stabilizes consistency training by keeping the condition $\mu$ frozen.

\noindent \textbf{Shared Dropout}
Although diffusion-based consistency models and rectified flow \cite{song2023improved, kim2025simple} proved that applying dropout improves the sample quality, the effect of dropout in consistency FM has not been investigated. Based on \cite{song2023improved}, while training the consistency FM model in the second stage, we apply dropout using the same random state for $v_{\theta}$ and $v_{\theta^{-}}$. Empirically, we use a dropout ratio of 0.05.

\noindent \textbf{Pseudo-Huber Loss}
Our loss in Eq. (\ref{eq:ms-cfm}) is composed of the $\ell_{2}$ metric. However, the $\ell_{2}$ metric imposes a large penalty on outliers, which can lead to an imbalanced loss with respect to time $t$, potentially increasing the gradient variance \cite{lee2024improving}. To mitigate this issue, we adopt the pseudo-Huber metric, which is less sensitive to outliers. We follow the implemented metric from \cite{song2023improved}, defined as $d(x,y)=\sqrt{||x-y||^{2}_{2}+c^{2}}-c$, where $c=0.00054\sqrt{d}$ and $d$ is the data dimension size.

\noindent \textbf{Delta Scheduling}
The term $\Delta t$ in Eq. (\ref{eq:ms-cfm}) is a parameter representing the time interval between two points on the trajectory, similar to the discretization size parameter in diffusion-based consistency models \cite{song2023consistency}. 
This implies that a small $\Delta t$ reduces the bias but increases the variance of the model, and vice versa for a large $\Delta t$. Since a small bias and large variance are desirable near the end of the training, inspired by \cite{song2023consistency}, we employ scheduling that progressively reduces $\Delta t$ during the training. 
To train the model uniformly across diverse $\Delta t$, we utilize linear step scheduling.
In practice, $\Delta t$ decreases from 0.1 to 0.001 over $K$ intervals, and each $\Delta t$ is used during $N/K$ epochs.

\noindent \textbf{Adversarial Learning}
To enhance synthesis quality, we utilize adversarial learning after the second stage. 
We adopt MSE-based adversarial loss $\mathcal{L}_{adv}$ \cite{mao2017least} and feature matching loss $\mathcal{L}_{fm}$ \cite{salimans2016improved} on the mel-spectrogram levels using a Conv2d discriminator \cite{li2022styletts}. Furthermore, we extend it to multi-segment adversarial learning for consistency FM. Let $x^{i}$ and $\hat{x}^{i}$ be the ground-truth and estimated endpoint at each segment, where each is obtained by $(i+1)/S \times x_{1} + (1-(i+1)/S)\times x_{0}$ and $f^{i}_{\theta}(t,x_{t}, \mu)$, respectively. 
Then, the objective function is defined as follows: 
\begin{equation}\label{eq:adv}
  \begin{gathered}
    \mathcal{L}_{adv}=||D(\hat{x}^{i})||^{2}_{2} + ||1-D(x^{i})||^{2}_{2}, \\
    \mathcal{L}_{fm} = \sum\nolimits_{l=1}^{L} ||D^{l}(\hat{x}^{i})-D^{l}(x^{i})||_{1}, \\
  \end{gathered}
\end{equation}
where $D$ indicates the discriminator; $D^{l}$ denotes the $l^{th}$ feature map of $D$. By applying multi-segment adversarial learning, we improve the endpoint quality of each segment, yielding better few-step performance. Finally, in this stage, we optimize $\mathcal{L}_{cfm}$, $\mathcal{L}_{adv}$, and $\mathcal{L}_{fm}$ with the ratio of 3, 1, and 2.

\section{Experiments}

\subsection{Experiment Setup}
\label{sec:experiment setup}

\noindent \textbf{Dataset} 
To verify RapFlow-TTS, we conducted experiments on the LJSpeech \cite{ljspeech17} and VCTK \cite{vctk2019} datasets. 
The LJSpeech is a single-speaker English dataset containing about 24 hours of audio clips. 
For the dataset split, we followed the settings of previous work in \cite{shen2018natural}. 
The VCTK is a multi-speaker English dataset composed of about 44 hours of speech recorded by 110 speakers. 
We randomly split the train, valid, and test sets in the ratios of 80\%, 10\%, and 10\%, respectively. To make the models compatible with the HiFiGAN vocoder used in the experiments, we followed the acoustic settings of HiFiGAN-V1 \cite{kong2020hifi}.

\noindent \textbf{Baselines} 
For comparison, we set the following systems as baselines: \textit{1) GT}, ground truth; \textit{2) VOC}, ground truth reconstructed by the vocoder; \textit{3) FastSpeech2} \cite{ren2020fastspeech}, Transformer-based TTS; \textit{4) Grad-TTS} \cite{popov2021grad}, diffusion-based TTS; \textit{5) Comospeech} \cite{ye2023comospeech}, consistency model with diffusion-based TTS; \textit{6) VoiceFlow} \cite{guo2024voiceflow}, rectified flow for TTS; and \textit{7) Matcha-TTS} \cite{mehta2024matcha}, flow matching for TTS. We recorded the performance of the baselines after training with their official codes. 
Since there is no official code for FastSpeech2, we used Meta's checkpoint.\footnote{\url{https://github.com/facebookresearch/fairseq}}

\noindent \textbf{Implementation Details}
For training, we set the number of epochs $N$ to 1400 and 1000 for the LJSpeech and VCTK datasets, respectively. 
We additionally used 150 and 50 epochs for adversarial learning, respectively. 
We used Adam optimizer with a learning rate of $1\times10^{-4}$ and batch size of 16. Regarding the hyperparameters of consistency FM, we set the multi-segment size $S$ to 2, the bin size $K$ for delta scheduling to 8, and the loss weight $\alpha$ in Eq. (\ref{eq:ms-cfm}) to $1\times10^{-5}$. The encoder freeze and shared dropout were applied as defaults, and the results with other techniques were shown through $\dagger$ notation in the experiments. In addition, we followed the network hyperparameters of Matcha-TTS. Finally, we applied an Euler solver to the ODE-based TTS models to synthesize speech. To investigate the few-step performance of ODE-based models, we set the number of function evaluations (NFE) to 2. To ensure reasonable performance for models that do not build a consistency model, such as Matcha-TTS and Grad-TTS, we also recorded their performances using NFE of 10 and 25, respectively.

\noindent \textbf{Evaluation Metrics} 
We assessed the results in terms of speech intelligibility, naturalness, and real-time factor (RTF). 
For intelligibility, we measured word error rate (WER) by applying the Whisper-M-based ASR model \cite{radford2023robust} to the synthesized speech. 
Regarding naturalness, we conducted a mean opinion score (MOS) test.
Twenty participants were asked to assess the naturalness of the synthesized speech by scoring it from 1 to 5 ranges. 
Twenty utterances were randomly selected from each system.
In addition, we used NISQA \cite{mittag2021nisqa}, a model-based naturalness assessment method, for ablation studies.
We provided evaluation results with a 95\% confidence interval.
Finally, we recorded the RTF, the ratio between the model synthesis time and duration of the synthesized speech, using a single Tesla V100 GPU to compare the inference speeds.

\subsection{Experimental Results}
\renewcommand{\thefootnote}{}
\footnotetext{$\dagger$ indicates the improved techniques are applied.}
\renewcommand{\thefootnote}{\arabic{footnote}} 
%

Table \ref{tab:main} shows the evaluation results of which findings can be analyzed as follows:
Firstly, the conventional score-based (Grad-TTS) and FM-based (VoiceFlow and Matcha-TTS) models exhibited significant degradation in naturalness (MOS) with the 2-step synthesis compared to the many-step synthesis.
By training with the consistency FM objective, the proposed RapFlow-TTS achieved superior performance with a 5- to 10-fold reduction in NFE than previous methods. In addition, it outperformed Comospeech in both speech intelligibility (WER) and naturalness. These results indicate that training consistency on a straightened ODE path leads to an effective consistency model, thereby improving few-step performance. However, the limited performance of the 2-step synthesis of the previous methods is attributed to the construction of consistency models on complex ODE paths (Comospeech) or the absence of consistency in FM-based methods (Matha-TTS and VoiceFlow). 
Secondly, the improved techniques in RapFlow-TTS$^{\dagger}$ contributed to significant enhancements in both the intelligibility and naturalness aspects. 
This is an encouraging result considering that the 2-step speed of RapFlow-TTS is comparable to the speed of FastSpeech2, which is generally adopted as a benchmark for fast TTS models. 
In summary, RapFlow-TTS synthesizes high-quality speech with fewer steps, achieved through consistency FM and improved training strategies, mitigating the inference speed limitations of previous ODE-based TTS.

\begin{table}
\caption{Evaluation results on the LJSpeech test set.}\label{tab:main}
\centering
\resizebox{1\linewidth}{!}{%
\begin{tabular}{lccccc}
\toprule
Model & Params & NFE & RTF & WER & MOS  \\
\midrule
GT          & - & - & -     & 3.31 $\pm$ 0.88 & 4.42 $\pm$ 0.08 \\
VOC         & - & - & 0.022 & 3.43 $\pm$ 0.95 & 4.30 $\pm$ 0.08 \\
\midrule
Grad-TTS    & 14.8M & 25 & 0.133 & 4.32 $\pm$ 0.82 & 3.78 $\pm$ 0.09 \\
VoiceFlow   & 14.8M & 10 & 0.069 & 4.04 $\pm$ 0.81 & 3.42 $\pm$ 0.09 \\
Matcha-TTS  & 18.2M & 10 & 0.056 & 3.28 $\pm$ 0.78 & 3.83 $\pm$ 0.09 \\
\midrule
\midrule
FastSpeech2 & 41.2M & 1  & 0.029 & 5.58 $\pm$ 1.04 & 3.35 $\pm$ 0.10 \\
Comospeech  & 14.8M & 2  & 0.034 & 5.41 $\pm$ 0.93 & 3.19 $\pm$ 0.10 \\
Grad-TTS    & 14.8M & 2  & 0.034 & 4.06 $\pm$ 0.79 & 1.92 $\pm$ 0.07 \\ 
VoiceFlow   & 14.8M & 2  & 0.034 & 4.00 $\pm$ 0.79 & 3.05 $\pm$ 0.09 \\
Matcha-TTS  & 18.2M & 2  & 0.031 & 3.15 $\pm$ 0.76 & 3.32 $\pm$ 0.10 \\
\midrule
RapFlow-TTS & 18.2M  & 2  & 0.031 & 3.33 $\pm$ 0.80 & 3.89 $\pm$ 0.09 \\
RapFlow-TTS$^{\dagger}$ & 18.2M  & 2  & 0.031 & \textbf{3.11} $\pm$ 0.76 & \textbf{4.01} $\pm$ 0.09  \\
\bottomrule
\end{tabular}}
\end{table}

\begin{table}
\centering
\caption{Ablation study results on the LJSpeech test set: All the samples in (A)-(H) models were synthesized with 2 NFE steps.}\label{tab:ablation}
\resizebox{1\linewidth}{!}{%
\begin{tabular}{lcc}
\toprule
Model & WER & NISQA  \\
\midrule
\textit{(A)}: Straight flow (Stage 1)                            & 3.19 $\pm$ 0.79 & 3.46 $\pm$ 0.04   \\
{\textit{(B)}: \textit{(A)} + Consistency FM}               & 3.51 $\pm$ 0.80 & 3.71 $\pm$ 0.05   \\
{\textit{(C)}: \textit{(B)} + Encoder freeze}               & 3.48 $\pm$ 0.80 & 3.73 $\pm$ 0.05   \\
{\textit{(D)}: \textit{(C)} + Shared dropout (RapFlow-TTS)} & 3.33 $\pm$ 0.80 & 3.78 $\pm$ 0.05 \\
\midrule
{\textit{(E)}: \textit{(D)} + Huber loss}                   & 3.42 $\pm$ 0.78 & 3.87 $\pm$ 0.04 \\
{\textit{(F)}: \textit{(D)} + Delta scheduling}             & 3.28 $\pm$ 0.78 & 3.90 $\pm$ 0.05 \\
\hspace{6pt}  linear $\rightarrow$ exp schedule             & 3.34 $\pm$ 0.79 & 3.78 $\pm$ 0.05 \\
{\textit{(G)}: \textit{(D)} + Adversarial learning}         & 3.24 $\pm$ 0.78 & 4.19 $\pm$ 0.04 \\
\midrule
{\textit{(H)}: Putting it together (RapFlow-TTS$^{\dagger}$)} & \textbf{3.11} $\pm$ 0.76 & 4.25 $\pm$ 0.04 \\
\hspace{6pt} w/ $NFE=10$            & 3.41 $\pm$ 0.74 & \textbf{4.28} $\pm$ 0.04 \\
\hspace{6pt} w/ $NFE=25$            & 3.47 $\pm$ 0.80 & \textbf{4.29} $\pm$ 0.04 \\
\bottomrule
\end{tabular}}
\end{table}

\subsection{Further Experiments}
\label{ssec:ablation}

\noindent \textbf{Ablation Studies}
In the first block of Table \ref{tab:ablation}, we analyzed the results of consistency FM and default techniques such as encoder freeze and shared dropout.
Experiment \textit{(A)} represents the results from stage 1, described in Section \ref{ssec:cfmtts}, where only a straight flow was trained without consistency.
Experiment \textit{(B)} presents the results of stage 2, in which the consistency FM was additionally trained at the stage-1 model. Despite some decrease in speech intelligibility, we observed a notable improvement in naturalness. It demonstrates that building a consistency model on a straight ODE trajectory is effective in enhancing the quality of the few-step synthesis. 
Meanwhile, experiment \textit{(C)} and \textit{(D)} show the results obtained using encoder freeze and shared dropout, respectively. Although the impact of the encoder freeze on the performance was not substantial, we empirically observed that it increased training speed by 1.3 times. Shared dropout improved the robustness of the consistency FM model, leading to enhanced performance across all metrics.

In the second block of Table \ref{tab:ablation}, we investigated the effectiveness of other improved techniques for RapFlow-TTS as follows: Firstly, in experiment \textit{(E)}, replacing the $\ell_{2}$ metric with the Huber metric improved the speech naturalness by reducing the gradient variance. Although we observed a slight decrease in WER performance, we adopted this strategy since we prioritized naturalness enhancement. Secondly, delta scheduling (experiment \textit{(F)}) improved both WER and NISQA. It indicates that the strategy, progressively reducing $\Delta t$ during training, is effective for performance improvement by minimizing the bias at the end of training. 
In addition, the performance degradation observed when using exponential scheduling suggests that linear scheduling, which uniformly trains various $\Delta t$ during training, is appropriate.
Lastly, in experiment \textit{(G)}, we validated the effects of adversarial learning. This strategy yielded the most significant improvements compared to previous techniques. By learning the ground-truth characteristics at the endpoints of each segment, Rapflow-TTS with adversarial learning can synthesize high-quality speech even in 2-step.

Finally, in experiment \textit{(H)}, where we combined all improved techniques, RapFlow-TTS achieved the best WER and NISQA. It suggests that each technique contributes to performance improvement as its unique method without redundancy. Furthermore, we utilized NFE of 10 and 25 for experiment \textit{(H)} to analyze the performance of RapFlow-TTS depending on NFE. We observed that naturalness slightly improved as NFE increased, while speech intelligibility deteriorated. This can be attributed to the characteristics of consistency models \cite{ye2023comospeech}, where performance improvements are minimal or even degrade in many-step synthesis. In future work, we further analyze this issue and explore potential improvements.


\noindent \textbf{Results on the multi-speaker dataset}
To verify RapFlow-TTS on the multi-speaker dataset, we conducted experiments on the VCTK dataset. Considering the increase in dataset size and number of speakers, we raised the shared dropout to 0.1 for model robustness. As depicted in Table \ref{tab:vctk}, RapFlow-TTS without improved techniques demonstrated lower speech intelligibility compared to the stage-1 model. 
However, in terms of naturalness, it significantly outperformed the stage-1 model in 2-step synthesis, even achieving a quality comparable to the stage-1 model's 10-step performance.
In addition, the improved techniques substantially enhanced both speech intelligibility and naturalness, enabling the synthesis of high-quality and intelligible speech in only 2 steps. These experimental results indicate that RapFlow-TTS can achieve an outstanding few-step synthesis performance even on multi-speaker datasets.

\begin{table}
\centering
\caption{Evaluation results on the VCTK test set.}\label{tab:vctk}
\resizebox{0.83\linewidth}{!}{%
\begin{tabular}{lccc}
\toprule
Model & NFE & WER &  MOS \\
\midrule
GT          & - & 1.79 $\pm$ 0.21 & 4.42 $\pm$ 0.07  \\
VOC         & - & 2.10 $\pm$ 0.23 & 4.06 $\pm$ 0.09  \\
\midrule
Straight flow (Stage 1)   & 2  & 2.01 $\pm$ 0.24 & 3.53 $\pm$ 0.09 \\
Straight flow (Stage 1)   & 10 & 2.52 $\pm$ 0.28 & 3.83 $\pm$ 0.09 \\
\midrule
RapFlow-TTS & 2 & 3.25 $\pm$ 0.31 & 3.83 $\pm$ 0.10 \\
RapFlow-TTS$^{\dagger}$ & 2 & \textbf{2.01} $\pm$ 0.23 & \textbf{4.28} $\pm$ 0.08 \\
\bottomrule
\end{tabular}}
\end{table}

\section{Conclusion}
In this study, we proposed RapFlow-TTS, the ODE-based TTS model capable of fast and high-fidelity speech synthesis. 
Based on consistency FM, RapFlow-TTS constructs a consistency model on a straight flow. 
By leveraging both the straight trajectory and the velocity-consistency property, high-quality speech synthesis was achievable in significantly fewer steps. 
In addition, training techniques such as time interval scheduling and adversarial learning further enhanced performance.
As a result, RapFlow-TTS outperformed existing ODE-based models and effectively overcame the trade-off between quality and speed.

\ifinterspeechfinal     
    \section{Acknowledgements}
     This research was supported by Voice team, NAVER Cloud, Republic of Korea. This research was supported by the BK21 FOUR funded by the Ministry of Education of Korea and National Research Foundation of Korea. This research was also results of a study on the ``Leaders in INdustry-university Cooperation 3.0" Project, supported by the Ministry of Education and National Research Foundation of Korea.     
\fi

\bibliographystyle{IEEEtran}
\bibliography{paper}

\end{document}